\documentstyle[twoside,12pt,epsf]{article}
\normalsize
\newcommand{\bdi}{\begin{displaymath}}
\newcommand{\edi}{\end{displaymath}}
\newcommand{\beq}{\begin{equation}}
\newcommand{\eeq}{\end{equation}}
\newcommand{\beqa}{\begin{eqnarray}}
\newcommand{\eeqa}{\end{eqnarray}}
\newcommand{\wt}{\widetilde}
\newcommand{\ra}{\rightarrow}

\begin{document}

\begin{titlepage}

\begin{flushright}
\today
\end{flushright}

\vspace{1cm}
\begin{center}
{\Large \bf Hopf instantons in Chern--Simons theory}\\[1cm]
C. Adam* \\
Institut f\"ur theoretische Physik, Universit\"at Karlsruhe \\

\medskip

\medskip

B. Muratori**,\, C. Nash*** \\
Department of Mathematical Physics, National University of Ireland, Maynooth
\vfill
{\bf Abstract} \\
\end{center}
We study an Abelian Chern--Simons \& Fermion system in three dimensions.
In the presence of a fixed prescribed background magnetic field we find
an infinite number of fully three-dimensional solutions. These solutions
are related to Hopf maps and are, therefore, labelled by the Hopf index.
Further we discuss the interpretation of the background field.

\vfill

$^*)${\footnotesize  
email address: adam@maths.tcd.ie, adam@pap.univie.ac.at} 

$^{**})${\footnotesize
email address: bmurator@fermi1.thphys.may.ie} 

$^{***})${\footnotesize
email address: cnash@stokes2.thphys.may.ie} 
\end{titlepage}

\section{Introduction}

It is the purpose of this letter to construct an infinite number of fully 
three-dimensional solutions to a Chern--Simons \& Fermion system that
are labelled by the Hopf index, thereby merging two features --
Chern--Simons theory and Hopf maps -- that have recently received wide 
attention within quantum field theory.

On the one hand, Hopf maps are just maps $S^3 \ra S^2$. These maps fall into
different homotopy classes that are labelled by the integers (the Hopf
index, see below for details). 
Field configurations with nontrivial Hopf index have already
been studied for some time in fluid dynamics \cite{KuMi}, in astrophysics
\cite{Giov,JaPi3}, in magnetic solids (magnetic solitons, see e.g., \cite{KIK})
and in classical electro-magnetism \cite{Ran1}. Recently, there has been
a lot of interest in field theories where static solutions with
nontrivial Hopf index (Hopf solitons) occur (see e.g., \cite{FN1}--\cite{AFZ}).

On the other hand, Chern--Simons theories with matter and/or a Maxwell term
have been studied intensively since their introduction \cite{DJT1,DJT2}.
When an Abelian Chern--Simons term in three dimensions is coupled to matter,
the magnetic field is forced to be proportional to the electric current
due to the equations of motion \cite{JLW}--\cite{Dun1}. 
Further, in these models there
exist soliton-like, static (i.e., two-dimensional) solutions that are
related to some topological invariants (e.g., maps $S^2 \ra S^2$) 
\cite{JLW}--\cite{Dun1}. 
Usually, these solitons behave like vortices, and, because of their 
topological nature, they exhibit magnetic flux quantization.
Therefore these solutions are physically relevant in situations
where the phenomenon of magnetic flux quantization occurs and where
matter is confined to a plane, the most prominent example being the
quantum Hall effect \cite{ZHK,Jain}.

At this point the question arises whether there exist fully
three-dimensional solutions for such Chern--Simons \& matter systems,
and whether these solutions may be characterized by some topological
invariants, as well.   

In this letter we shall demonstrate that, if the presence of a 
fixed, prescribed 
background magnetic field is assumed, then there indeed exist solutions to the
Chern--Simons \& Fermion system defined below. Further, these solutions are 
related to Hopf maps and are, therefore, labelled by the Hopf index.
Hence we find, on one hand, a three-dimensional version of the
phenomenon of magnetic flux quantization, namely topologically quantized
magnetic knots (Hopf maps are related to magnetic knots, see
\cite{Ran1,RaTr1}). On the other hand, a Chern--Simons term is
well-known to be induced in QED$_3$ (see e.g. 
\cite{DJT2,Red1,Alva1,Poly1,DW1}). 
Further, a
Chern--Simons term is used for topological generation of mass in some
three-dimensional QFTs \cite{DJT1,DJT2}. Whenever Fermions are included in
such theories, our solutions (the Hopf instantons) should be relevant
for the study of non-perturbative features of these QFTs (as well as
of QED$_3$).

In the main section of this letter we define the model and construct  
solutions to its equations of motion, as well as the Hopf maps that provide 
these solutions. In the final section we give two possible interpretations
for the background field that is present in our solutions.

\section{Construction of the Hopf instantons}

We start with the action ($i,j,k =1 \ldots 3$)
\beq
S=\int d^3 x \Bigl( \Psi^\dagger (-i\partial_j -\bar A_j)\sigma_j \Psi
+ \frac{1}{2}\vec A \vec B \Bigr)
\eeq
where $\Psi$ is a two-component spinor (Fermion), $\sigma_j$ are the
Pauli matrices and $\vec A$ is an Abelian gauge potential. Further,
\beq
S_{\rm CS}=
\frac{1}{2}\int d^3 x\vec A\vec B =\frac{1}{4}\int d^3 x 
\epsilon_{ijk}A_iF_{jk}
\eeq
is the Chern--Simons (CS) action, where the Chern--Simons coupling 
constant is chosen equal to one; and
\beq
\bar A_i = A_i + A_i^{\rm B}
\eeq
where the background gauge field $A_i^{\rm B}$ and its magnetic field
$B_i^{\rm B}=\epsilon_{ijk}\partial_j A^{\rm B}_k$ are
\beq
\vec A^{\rm B} 
=-\frac{1}{1+r^2} \vec N \, , \quad 
\vec B^{\rm B} =- \frac{4}{(1+r^2)^2} \vec N
\eeq
and we have introduced the unit vector
\beq
\vec N = \frac{1}{1+r^2}
\left( \begin{array}{c} 2x_1 x_3 -2x_2  \\ 2x_2 x_3 +2x_1 \\
1-x_1^2 -x^2_2 +x_3^2 \end{array} \right) 
\eeq
($\vec N^2 =1$) for later convenience.
Observe that the (fixed, non-dynamical) background field is coupled to the
Fermion, but it is absent in the CS term. The equations of motion resulting
from the action (1) are
\beq
(-i\partial_j -\bar A_j )\sigma_j \Psi =0,
\eeq
(the Dirac equation) and
\beq
\vec\Sigma := \Psi^\dagger \vec\sigma \Psi = \vec B .
\eeq
Observe that for any pair $(\Psi ,\bar A_j)$ that solves the Dirac equation (6)
the spin density $\vec\Sigma$ related to $\Psi$ has to obey
 \beq
\vec\partial \vec\Sigma =0,
\eeq
therefore, equations (6) and (7) are consistent.

The simplest solution to this system is (see \cite{LoYa,Zero})
\beq
\Psi =\frac{4}{(1+r^2)^\frac{3}{2}} ({\bf 1} +i\vec x \vec\sigma )
\left( \begin{array}{c} 1 \\ 0 \end{array} \right) 
\eeq
\beq
\vec A 
=\frac{4}{1+r^2} \vec N  
\eeq
\beq
\vec\Sigma =\vec B = \frac{16}{(1+r^2)^2} \vec N
\eeq
($\vec N$ is given in (5)).
Here the dynamical gauge field is proportional to the background field,
therefore one could find a solution without background field by choosing
either a different normalization of the fermion (9) or by choosing a
Chern--Simons coupling constant in (2), (7) different from 1.
However, this will not be true for the solutions below, for which the
background field (4) is crucial.

 Concerning the geometrical behaviour of the magnetic field $\vec B$ in
(11), it is important to note that it is related to a Hopf map, and,
therefore, the value of the CS action (2) is topologically quantized by
the Hopf index. This we want to explain now.

Generally, a complex function $\chi : {\rm\bf R}^3 \ra {\rm\bf C}$ with
the additional property $\lim_{r\to\infty} \chi =\chi_0 = {\rm const.}$
defines a Hopf map $\chi :S^3 \ra S^2$, where the coordinates in ${\rm\bf
R}^3$ and ${\rm\bf C}$ are stereographic coordinates of the $S^3$ and
$S^2$, respectively. The pre-images in ${\rm\bf R}^3$ of points
$\chi = {\rm const.}$ are closed curves in ${\rm\bf R}^3$ (circles in
$S^3$), and any two different circles are linked exactly $N$ times, where
$N$ is the Hopf index. Further, a magnetic field $\vec B$ 
(the Hopf curvature) is related to the
Hopf map $\chi$ via
\beq
\vec B = \frac{2}{i}\frac{(\vec\partial\bar\chi)\times(\vec\partial\chi)}{
(1+\bar\chi \chi)^2} =4\frac{S(\vec\partial S)\times\vec
\partial\sigma}{(1+S^2)^2}
\eeq
where $\chi =Se^{i\sigma}$ is expressed in terms of its modulus $S$
and phase $\sigma$ at the r.h.s. of (12).

Mathematically, the curvature $F=\frac{1}{2}F_{ij}dx_i dx_j$, $F_{ij}=
\epsilon_{ijk}B_k$, is the pullback under the Hopf map, $F=\chi^* \Omega$,
of the standard area two-form $\Omega$ on the target $S^2$ 
with radius 1 (in stereographic
coordinates) \cite{Ran1,FN1,BS1}, 
\beq
\Omega \equiv \frac{2}{i}\frac{d\bar z dz}{(1+z\bar z)^2} .
\eeq
Geometrically, $\vec B$ is tangent to the closed curves $\chi ={\rm const}$.
The Hopf index $N$ of $\chi$ may be computed from $\vec B$ via
\beq
N=\frac{1}{16\pi^2}\int d^3 x \vec A \vec B
\eeq
where $\vec B=\vec\partial\times \vec A$.

The simplest (standard) Hopf map $\chi^{(1)}$ with Hopf index $N=1$ is
\beq
\chi^{(1)}=\frac{2(x_1 +ix_2)}{2x_3 -i(1-r^2)}
\eeq
(a Hopf map has to be single valued, but may well be singular, as $\chi 
=\infty$ is just the south pole of the target $S^2$). 
This $\chi^{(1)}$, (15), leads to a Hopf curvature $\vec B^{(1)}$
via (12) that is just the magnetic field (11). Here the question arises
whether there are more solutions to (6), (7) that are characterized by
higher Hopf maps, and we will find that this is indeed the case.

[Remark: there is some arbitrariness in the choice of the normalization factor 
in front of the area two-form (13). We chose it for a target two-sphere
of radius one. Consequently, the magnetic field (11) is the Hopf
curvature of the
standard Hopf map (15) with Hopf index 1, and the background field (4) 
is minus 1/4 the Hopf curvature of the standard map (15).
There are other choices in the literature, e.g. for target two-spheres
with radius $1/\sqrt{2}$ or 1/2. E.g., for radius $1/\sqrt{2}$ the magnetic
field (11) is twice the Hopf curvature of the standard Hopf map (15) (and the
background field (4) is minus 1/2 the Hopf curvature of the standard map (15)).
Such higher (integer)
multiples of $\vec B^{(1)}$ may be expressed by higher
Hopf maps as follows.
Rewrite $\chi^{(1)}$ as $\chi^{(1)} =S\exp (i\sigma )$, where $S$ and
$\sigma$ are the modulus and phase of $\chi^{(1)}$ respectively. 
Then $n\vec B^{(1)}$
may be computed via (12) from $\chi^{(1)}_n=S\exp (in\sigma)$. Here
only integer $n$ are allowed, because for a Hopf map with nonzero Hopf index 
the phase $\sigma$ necessarily is multiply valued (see \cite{Ran1} for
details). It follows that the background field (4) is not a Hopf curvature
(but a fraction thereof) in our choice of conventions.]

Next we should provide some more Hopf maps that will give rise to more
solutions to (6), (7).
We will produce these Hopf maps by composing the
standard Hopf map with some maps $S^2 \ra S^2$, i.e.,
\beq
\chi_R :S^3 \stackrel{\chi^{(1)}}{\ra} S^2 \stackrel{R}{\ra} S^2 .
\eeq
In fact, as we use stereographic coordinates on $S^2$, we will use rational 
maps for $R$,
\beq
R: z \ra R(z)=\frac{P(z)}{Q(z)}
\eeq
which are well-known to produce maps $S^2 \ra S^2$ (here $P$, $Q$ are
polynomials) \cite{HMS,Horv1}. 
The winding number of the map is equal to the degree of
the rational map, ${\rm deg}(R)$,
\beq
{\rm deg}(R)={\rm max}(p,q)
\eeq
where $p$ and $q$ are the degrees of the polynomials $P$ and $Q$. In fact,
we want to restrict to the simplest maps
\beq
R_n (z)=z^n
\eeq
i.e., we map
\beq
\chi^{(n)} = (\chi^{(1)})^n
\eeq
or, for modulus $S=:T^{1/2}$ and phase $\sigma$ of $\chi^{(1)}$,
\beq
(S,\sigma) \ra (S^n, n\sigma ),
\eeq
where the modulus and phase of the standard Hopf map (15) are
\bdi
T:=S^2 = \frac{4(r^2 -x_3^2)}{4x_3^2 +(1-r^2)^2}
\edi
\beq
\sigma = \sigma^{(1)} + \sigma^{(2)} \, \quad 
\sigma^{(1)} =\arctan \, \frac{x_2}{x_1} \sigma^{(2)} = \arctan \, 
\frac{1-r^2}{2x_3} .
\eeq
Hopf maps that are composed like (16) lead to a Hopf index $N=n^2$, where 
$n$ is the winding number (i.e., degree) of the rational map
(see e.g. \cite{RaTr1}, where these Hopf maps have been discussed).

For the Hopf curvature this implies 
\beq
B^{(n)}_i =2\epsilon_{ijk}\frac{(T^n)_{,j}n\sigma_{,k}}{(1+T^n)^2}=
n^2\frac{T^{n-1}(1+T)^2}{(1+T^n)^2}B^{(1)}_i ,
\eeq
where $B^{(1)}_i $ is just the standard Hopf curvature (11). The logarithm
of the factor in front
of $B^{(1)}_i$ at the r.h.s. of (23), when viewed as a function of $T$,
\beq
M_n (T) \equiv \ln n^2\frac{T^{n-1}(1+T)^2}{(1+T^n)^2}
\eeq
obeys the following nonlinear differential equation ($'\equiv \partial /
\partial T $)
\beq
M' +TM'' =-2\frac{e^M -1}{(1+T)^2} 
\eeq
for all $n$, as may be checked easily.

Next we need some facts about the Dirac equation (6). Suppose a pair
$(\Psi ,\bar A_i)$ is given that solves (6), then $\bar A_i$ may be 
expressed in terms of the zero mode $\Psi$ as
\bdi
\bar A_i =\frac{1}{|\vec \Sigma |}(\frac{1}{2}\epsilon_{ijk}\partial_j
\Sigma_k +{\rm Im}\, \Psi^\dagger \partial_i \Psi )   
\edi
\beq 
= \frac{1}{2}\epsilon_{ijk}(\partial_j \ln |\vec \Sigma |){\cal N}_k
+\frac{1}{2}\epsilon_{ijk}\partial_j {\cal N}_k +
{\rm Im}\, \widehat\Psi^\dagger
\partial_i \widehat\Psi
\eeq
where we have expressed $\bar A_i$ in terms of the general
unit vector and unit spinor
\beq
\vec {\cal N}=\frac{\vec \Sigma}{|\vec \Sigma |}\, ,\quad \widehat\Psi =
\frac{\Psi}{|\Psi^\dagger \Psi |^{1/2}} .
\eeq
Now we introduce the zero modes 
\beq
\Psi^{(M)}=e^{i\Lambda}e^{M/2}\Psi
\eeq
where $\Psi$ is the zero mode (9) and $M$ is an (at the moment arbitrary)
function of $T$. The pure gauge factor $\Lambda$ will be determined
accordingly below. Due to the fact that
$T_{,i}\Sigma_i \equiv T_{,i}B^{(1)}_i =0$
(where $B^{(1)}_i$ is given in (11)), which is obvious from (12), it is
still true that $\Sigma^{(M)}_{i,i}=0$, i.e., $\Psi^{(M)}$ really is
a zero mode. The corresponding gauge field $\bar A^{(M)}_i$ that 
solves the Dirac equation together with $\Psi^{(M)}$ reads (we use (26))
\beq
\bar A^{(M)}_i =\bar A^{(0)}_i +\frac{1}{2}\epsilon_{ijk}(\partial_j M)
N_k +\Lambda_{,i}
\eeq
where $\bar A^{(0)}_i$ is the gauge field (10) plus the background gauge
field (4), and $\vec N$ is the specific unit vector (5).
For the corresponding magnetic field $\bar B^{(M)}_i$ we find
\beq
\bar B^{(M)}_l =\bar B^{(0)}_l +\frac{1}{2}[M' (T_{,lk}N_k +T_{,l}N_{k,k}
-T_{,kk}N_l - T_{,k}N_{l,k})-M'' (T_{,k})^2 N_l]
\eeq
where we have used $M_{,k}=M' T_{,k}$, $T_{,k}N_k =0$ and $(\vec N^2)_{,k}=0$.
After some tedious algebra, we arrive at the expression
\beq
\bar B^{(M)}_l =\bar B^{(0)}_l -\frac{8(1+T)^2}{(1+r^2)^2}(M' +TM'' )N_l
\eeq
Now we want to insert this into the second equation, (7) (the Chern--Simons
equation). Therefore, we have to subtract the background field (4),
$\bar B^{(0)}_i - B^{\rm B}_i = B^{(1)}_i = 16(1+r^2)^{-2}N_i$. We find that
\beq
\bar B^{(M)}_l -B^{\rm B}_l = \frac{16}{(1+r^2)^2}N_l -
\frac{8(1+T)^2}{(1+r^2)^2}(M' +TM'' )N_l \stackrel{!}{=}\Sigma^{(M)}_l
=\frac{16e^M}{(1+r^2)^2} N_l
\eeq
or, after multiplication by $N_l$, 
precisely eq. (25). Hence,
we have shown that, in the presence of the fixed prescribed background
magnetic field (4), there exists an infinite number of fully three-dimensional
solutions to the system of equations (6), (7). 

Here we still should explain why $n$ is restricted to integer values,
which is related to the pure gauge factor $\Lambda$ in (28). The problem
is that the gauge potential (29) without the pure gauge term is singular.
For the explicit expressions $M_n$, (24), the gauge potential (29) may be 
rewritten as
\beq
A^{(M_n)}_j = A_j -\frac{(n-1)(1-T^{n+1}) +(n+1)(T-T^n )}{(1+T)(1+T^n)}
\sigma_{,j} +(n-1) \Lambda_{,j}
\eeq
where
\beq
\Lambda = \sigma^{(1)} - \sigma^{(2)}
\eeq
is chosen such that (33) is regular everywhere ($\sigma^{(1)} $ and
$\sigma^{(2)}$ are defined in (22)). This implies that $n$ has to be 
integer, because only for integer $n$ $\exp (i(n-1)\Lambda )$ (and
consequently the spinor (28)) will be single-valued. Further, we
may compute the resulting
Chern--Simons density 
\beq
\vec A^{(M_n)}\cdot \vec B^{(M_n)}=\frac{64 n^3}{(1+r^2)^3}
\frac{T^{n-1}(1+T)^2}{(1+T^n)^2}
\eeq
which, when expressed in spherical polar coordinates $(r,\vartheta
,\varphi)$ depends on $r,\vartheta$ only. 
By integrating (35) we can explicitly verify the relation $N=n^2$ for the
Hopf index $N$. We have not succeeded in integrating (35) analytically so
far, but (35) may easily be integrated in $r,\vartheta ,\varphi$
coordinates numerically with the help of mathematica. The integrand
is so well behaved that the numerical integration reproduces the integer
result $N=n^2$ without showing even a small numerical deviation. 
The only technicality is that
for large $n$ one has to subdivide the range of integration for
$\vartheta \in [0,\pi ]$, because the integrand (35) becomes rather
oscillatory for large $n$.

\section{Discussion}

We have shown that, in the presence of the fixed prescribed background
magnetic field (4), there exists an infinite number of fully three-dimensional
solutions to the system of equations (6), (7). Further, these solutions
(i.e., the magnetic fields) are the Hopf curvatures of the Hopf maps
(16), (20) and are, therefore, labelled by the corresponding Hopf index
$N=n^2 , \, n\in {\rm\bf Z}$.

Here the background field (4) was crucial, because it determines the 
inhomogeneous part of the non-linear differential equation (25), and,
because of this non-linearity, the inhomogeneous part of (25) 
crucially affects the nature of the solutions of (25). Differently 
stated, the Hopf instanton solutions (24) uniquely determine the background
field (4).

Before closing, we want to briefly discuss whether it is possible to
further interpret the background field (4), beyond just stating that its
presence is crucial for the existence of the solutions (16), (20).
Indeed, there are (at least) two interpretations that we want to
describe now.

Firstly, let us study the following zero modes $\Psi_l =\exp 
(i\Lambda +
M_l /2) \Psi$, analogous to (28), where we now choose
\beq
M_l = l\ln \frac{T}{1+T} 
\eeq
(and $\Lambda =l\varphi =l\sigma^{(1)}$ to achieve a non-singular
gauge potential for $M_l$).
It follows easily that 
\beq
\Psi_l =\frac{2^{l+2}r^l}{(1+r^2)^l}
({\bf 1} +i\vec x\vec \sigma )
\left( \begin{array}{c} Y_{l,l}(\vartheta ,\varphi) \\ 0 \end{array} \right) 
\eeq
where $Y_{l,l}$ are spherical harmonics. In other words, 
$\Psi_l$ is just a higher
angular momentum zero mode (with magnetic quantum number $m=l+1/2$) that
can be constructed from the simplest zero mode (9), see \cite{LoYa,Zero}.
The gauge field $\bar A^{(l)}_j $ that solves the Dirac equation (6) together 
with $\Psi_l$ may be computed easily, and its magnetic field is
\beq
\bar B^{(l)}_j =\frac{12+8l}{(1+r^2)^2}N_j
\eeq
i.e., adding one unit of angular momentum to $\Psi_l$ changes the
corresponding magnetic field by $8(1+r^2)^{-2}\vec N$, which is precisely 
minus two times the background magnetic field (4). It is, therefore, 
tempting to conjecture that the background field (4) is somehow related to the
half-integer intrinsic angular momentum (i.e., spin) of the fermion.
Of course, this is just an observation at this point, because a mechanism
that generates this background field is still missing.

Secondly, it is possible to re-interpret the background field $\vec A^{\rm B}$,
(4), as a spin connection $\omega$ in the Dirac equation (6) on a
conformally flat manifold with torsion. Generally, the Dirac operator with
spin connection reads (see e.g. \cite{Bertl} for details)
\beq
{\cal D} = \gamma^a E_a{}^\mu (\partial_\mu +A_\mu +\frac{1}{4}
[\gamma_b ,\gamma_c ]\omega^{bc}{}_\mu )
\eeq
where $\gamma^a$ ($\equiv \sigma^a$ in our case) are the usual Dirac matrices,
$E_a{}^\mu$ is the inverse vielbein and $\omega^{bc}{}_\mu$ is the spin
connection (here $\mu ,\nu$ are Einstein (i.e., space time) indices and
$a,b,c$ are Lorentz indices). Our Dirac equation (6) may be rewritten 
in the form of eq.  
(39) provided that the vielbein is conformally flat, $E_a{}^\mu =f\delta_a^\mu
$, where $f$ is an arbitrary function. Using $[\sigma_b ,\sigma_c ]=
2i\epsilon_{bcd}\sigma^d$ we find
\beq
\frac{i}{2}\delta_a^k \epsilon_{bcd}\sigma^a \sigma^d \omega^{bc}{}_k
\stackrel{!}{=} \delta_a^k \sigma^a A^{\rm B}_k
\eeq
(here $k$ is an Einstein index in three dimensions). The l.h.s. of (40) has
to be antisymmmetric in $a,d$, i.e., the quantity $\wt \omega_{da}
:= \delta_a^k \epsilon_{bcd}\omega^{bc}{}_k$ obeys $\wt \omega_{da}
=-\wt\omega_{ad}$. This leads to $\wt \omega_{ab} =\epsilon_{abc}
\delta_c^k A^{\rm B}_k$. If we further assume $\omega^{ab}{}_k =
-\omega^{ba}{}_k$ (i.e., covariant constancy of the metric) then we find that
\beq
\omega_{abk} =\delta_{ka}A^{\rm B}_b - \delta_{kb}A^{\rm B}_a
\eeq
(where $A^{\rm B}_a \equiv \delta_a^k A^{\rm B}_k$, i.e., it is {\em not}
the Lorentz vector $E_a{}^k A^{\rm B}_k$). Finally, we find for the torsion
$T$ (expressed in Lorentz indices only)
\beq
2T_{abc}= (\delta_{ab}\delta_c^k - \delta_{ac}\delta_b^k )\partial_k f
- (\omega_{abc} - \omega_{acb})
\eeq
where
\beq
\omega_{abc}=E_c{}^k\omega_{abk}=f\delta_c^k \omega_{abk} .
\eeq
Hence, with $\omega_{abk}$ given by (41), we may freely choose a conformally 
flat metric (i.e., conformal factor $f$) and compute the resulting torsion 
via (42). Due to the form of $\omega_{abk}$ (i.e., $\vec A^{\rm B}$) it is,
however, not possible to choose a conformal factor such that the torsion is
zero. On the other hand, it is possible to choose the flat metric $f=1$,
so that (the anti-symmetric part of) the spin connection is given just by 
the torsion. 

\section{Acknowledgments}
The authors thank M. Fry for helpful discussions. In addition,
CA gratefully acknowledges useful conversations with R. Jackiw.
CA is supported by a Forbairt Basic Research Grant.
BM gratefully acknowledges financial support from the Training and 
Mobility of Researchers scheme (TMR no. ERBFMBICT983476).

\end{document}